\documentclass[prd,twocolumn,floatfix,amsmath,nofootinbib,amssymb,floatfix]{revtex4}
\usepackage{graphicx,color,dcolumn,booktabs,bm}
\usepackage{longtable,lscape}
\usepackage{txfonts}
\usepackage{overpic}
\usepackage{amssymb}
\usepackage{indentfirst}
\usepackage{feynmf}   
\usepackage{slashed}  
\usepackage{cases}
\usepackage{color}
\usepackage{multirow}
\usepackage{slashed}
\usepackage{epstopdf}
\usepackage{longtable}
\usepackage{ulem}
\usepackage{graphicx,color,dcolumn,booktabs,bm}
\usepackage[colorlinks,
            citecolor=blue,
            anchorcolor=red,
            menucolor=red,
            linkcolor=red,
            filecolor=red,
            runcolor=red,
            urlcolor=blue,
            frenchlinks=red]{hyperref}

\graphicspath{{Figures/}} %
\allowdisplaybreaks

\begin{document}

\title{Potential higher radial excitations in the light pseudoscalar meson family}

\author{Li-Ming Wang$^{1,2}$}\email{lmwang15@lzu.edu.cn}
\author{Qin-Song Zhou$^{1,2}$}\email{zhouqs13@lzu.edu.cn}
\author{Cheng-Qun Pang$^{3,4}$}\email{pcq@qhnu.edu.cn}
\author{Xiang Liu$^{1,2,3}$\footnote{Corresponding author}}\email{xiangliu@lzu.edu.cn}
 \affiliation{
$^1$School of Physical Science and Technology, Lanzhou University, Lanzhou 730000, China\\
$^2$Research Center for Hadron and CSR Physics, Lanzhou University and Institute of Modern Physics of CAS, Lanzhou 730000, China\\
$^3$Joint Research Center for Physics, Lanzhou University and Qinghai Normal University, Xining 810000, China\\
 $^4$College of Physics and Electronic Information Engineering, Qinghai Normal University, Xining 810000, China}
\begin{abstract}
Inspired by the event accumulation around 2.6 GeV in the $\eta^\prime\pi^+\pi^-$ invariant mass spectrum of $J/\psi\to \gamma \eta^\prime\pi^+\pi^-$, which was reported by the BESIII Collaboration, we carry out the study of the mass spectrum and decay behavior of four radial excitations in the pseudoscalar meson family, which include $\eta^{(\prime)}(6S)$ and $\eta^{(\prime)}(7S)$. Combining with these analysis, we present the calculation of the reactions induced by a pion or kaon on the proton target which are relevant to these four discussed states. According to this information, we give concrete experimental suggestion of searching for them, which will become a new task for future experiments.
\end{abstract}
\date{\today}
\maketitle

\section{Introduction}\label{sec1}

Studying hadron spectroscopy can provide valuable hints to deep our understanding of nonperturbative quantum chromodynamics (QCD). Among the hadron family, light mesons construct a special group since there exists extra abundant measured information of light mesons. In 1935, Yukawa predicted the existence of the pion meson for quantitatively depicting a nuclear force \cite{Yukawa:1935xg}, which was discovered by Lattes {\it et al.} \cite{Lattes:1947mx}. By joint effort from experimentalist and theorist, the number of light mesons has been increasing in the past decades. In recent years, the running experiment like BESIII is still playing crucial role to explore and discover the light meson. As emphasized in a paper recently released by the BESIII Collaboration \cite{Ablikim:2019hff}, investigating the light hadron spectrum will be an important issue in the next ten years at BESIII. Here, besides establishing a conventional light meson, what is more important is the search for an exotic hadronic state like the glueball \cite{Klempt:2007cp}.

Since 2011, the Lanzhou group has performed systematic studies on mass spectrum and strong decay behavior of light mesons, which include pseudoscalar states \cite{Yu:2011ta,Wang:2017iai}, the vector state \cite{Wang:2012wa}, $\rho$ and $\rho_3$ states \cite{He:2013ttg}, tensor states \cite{Ye:2012gu}, pseudotensor states \cite{Wang:2014sea}, axial vector states \cite{Chen:2015iqa}, $J^{PC}=2^-$ unflavored states \cite{Guo:2019wpx}, kaons \cite{Pang:2017dlw}, and high-spin states \cite{Pang:2015eha}. Besides the above work, other theoretical groups focused on this issue \cite{Piotrowska:2017rgt}. By these investigations, an overview of the properties of the light meson family was presented, since the above studies almost contain these light mesons with allowed different spin-parity quantum numbers.

After finishing these theoretical works, we should consider how to continue the following studies of the light meson. There should exist three possible directions:
1) Combining with concrete experimental observation of light meson, we can decode the properties of these observed states by giving mass spectrum analysis, decay behavior, and production. A typical example is Ref. \cite{Wang:2019qyy}, where $X(2100)$ observed in $J/\psi\to \phi\eta\eta^\prime$ can be categorized into the $h_1$ meson group, which can be supported by mass and decay calculation. In addition, its production induced by pion and kaon was discussed.
2) In exploring high spin states, if checking the data from Particle Data Group (PDG) \cite{Tanabashi:2018oca}, we may find that the experimental information of higher-spin states is not abundant. When establishing these high-spin states in experiment, theoretical guidance is important. In Ref. \cite{Pang:2015eha}, the authors presented a systematic study on these high-spin states below 3 GeV. Additionally, Pang {\it et al.} predicted the mass and decay behavior of $5^{++}$ mesons \cite{Pang:2018gcn}.
3) Exploring the potential higher radial excitations of light meson accessible at experiment is also very interesting, and will be the main task in this work.

 As the frontier of precision in particle physics,
 hadron physics has entered a new era with the accumulation of high-precision experimental data since 2003. We may find two typical examples to reflect the importance of experimental precision. One is hidden-charm $P_c$ observation. In 2015,
LHCb observed two $P_c$ states [$P_c(4380)$ and $P_c(4450)$] in $\Lambda_b\to J/\psi p K$ \cite{Aaij:2015tga}. After four years, LHCb reanalyzed this channel based on data from Run I and Run II and found
that $P_c(4450)$ contains two substructures [$P_c(4440)$ and $P_c(4457)$] and there also exists another $P_c(4312)$ \cite{Aaij:2019vzc}, which give strong evidence of the existence of hidden-charm molecular pentaquark \cite{Chen:2019asm}.

\begin{figure}[h]
\centering
\begin{tabular}{cc}
\hspace{-20pt}
\includegraphics[width=0.45\textwidth,scale=0.35]{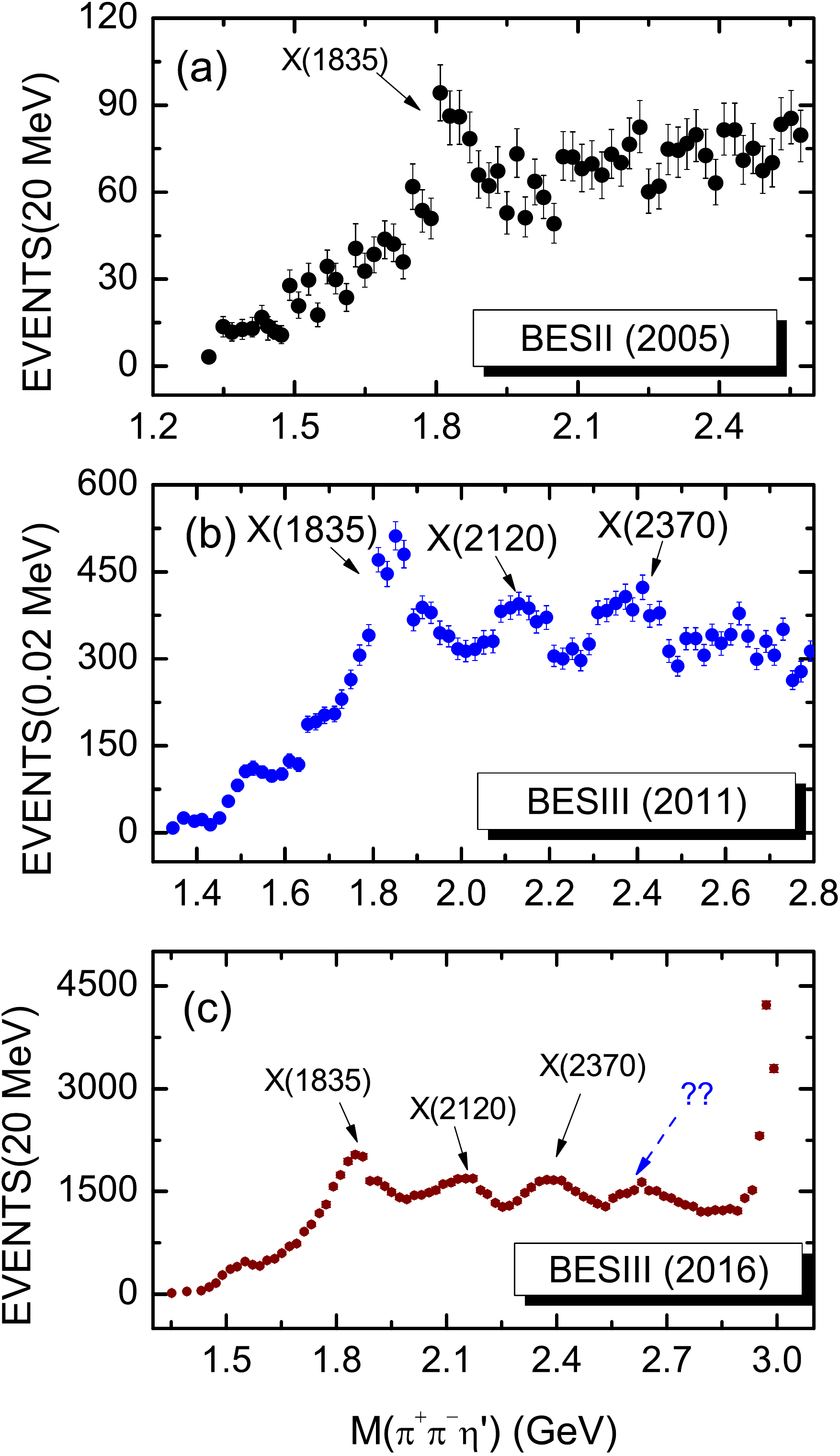}
\end{tabular}
\caption{The comparison of three experimental data of the $\eta^\prime \pi^+\pi^-$ invariant mass spectrum of $J/\psi\to \gamma\eta^\prime \pi^+\pi^-$  \cite{Ablikim:2005um,Ablikim:2010au,Ablikim:2016itz}.}\label{exp}
\end{figure}

Another example is about a series of observations of pseudoscalar $X$ states after $X(1835)$ as shown in Fig. \ref{exp}. In 2005, the BESII Collaboration reported $X(1835)$ in $J/\psi\to \gamma\eta^\prime \pi^+\pi^-$ \cite{Ablikim:2005um}. And then, BESIII not only confirmed $X(1835)$ but also found two structures $X(2120)$ and $X(2370)$ \cite{Ablikim:2010au} by analyzing the same decay process. With accumulation of experimental data, BESIII again studied the $J/\psi\to \gamma\eta^\prime \pi^+\pi^-$ channel \cite{Ablikim:2016itz}. And then, $X(2370)$ and $X(2500)$ were reported by BESIII in analyzing the $J/\psi\to\gamma\eta'\pi^+\pi^-$ \cite{Ablikim:2016itz} and $J/\psi\to\gamma\phi\phi$ \cite{Ablikim:2016hlu} decays, respectively.
Stimulated by these BESIII observations, the Lanzhou group found two Regge trajectories \cite{Yu:2011ta,Wang:2017iai}, which are composed of these $\eta$ mesons listed by the PDG \cite{Tanabashi:2018oca}, and these observed $X(1835)$, $X(2120)$, $X(2370)$
and $X(2500)$. This study \cite{Yu:2011ta,Wang:2017iai} provides valuable information when constructing the isoscalar pseudoscalar $\eta/\eta^\prime$ meson family.

We noticed that BESIII already collected a sample of 10 billion $J/\psi$ events in 2019. With more precision data, we have reason to believe that more $\eta/\eta^\prime$-like states with higher mass can be found in future experiments. Thus, it is a suitable time to carry out the theoretical study of the properties of these potential higher radial excitations in the light isoscalar pseudoscalar meson family.

Another topic of this work is to discuss the possible search for them by combing with the present BESIII experimental data  \cite{Ablikim:2005um,Ablikim:2010au,Ablikim:2016itz}. However, we have to face the  fact that the masses of $\eta(7S)$ and $\eta^\prime(7S)$ are close to that of $\eta_c(1S)$, which results in
the difficulty of detecting the signals of $\eta(7S)$ and $\eta^\prime(7S)$.
 Thus, we propose
that the production processes induced by a pion or kaon can be as a good platform to explore  higher states in the $\eta/\eta^\prime$ meson family. In this work, we take the production processes relevant to $\eta^{(\prime)}(6S)$ and $\eta^{(\prime)}(7S)$ as example to give a quantitative illustration.

The paper is organized as follows. After the Introduction, we present the mass spectrum analysis in Sec. \ref{sec2}. In Sec. \ref{sec3}, we calculate the Okubo-Zweig-Iizuka (OZI) allowed decays of these discussed $\eta/\eta^\prime$ states.
And then, we explore the production processes of four discussed $\eta/\eta^\prime$ states which are induced by the pion and kaon  (see Sec. \ref{sec4}). This paper ends with a summary.

\section{Estimating the mass of higher radial excitation of $\eta$ family}\label{sec2}

In Refs. \cite{Yu:2011ta,Wang:2012wa}, the Lanzhou group found two Regge trajectories for the $\eta$ and $\eta^{\prime}$ states, which are [$\eta(548)$, $\eta(1295)$, $\eta(1760)$, $X(2100)/X(2120)$, $X(2370)$] and [$\eta^\prime(958)$, $\eta(1475)$, $X(1835)$, $
\eta(2225)$, $X(2500)$]\footnote{We should indicate that the spin-parity quantum numbers of $X(2120)$, $X(2370)$, and $X(2500)$ have not been confirmed in experiment. In this work, we treat them as pseudoscalar states.}. In this section, we continue to estimate the masses of the fifth and the sixth radial excited states of the pseudoscalar meson family, which is an extension of the above two Regge trajectories. 

The Regge trajectory theory was first proposed by Regge in 1959, and was later widely used to study the light hadron spectrum \cite{Yu:2011ta,Wang:2012wa,Ye:2012gu,He:2013ttg,Ebert:2009ub,Gauron:2000ri,Anisovich:2000kxa,Ebert:2009ua}.
In general, for a light meson, the relationship between the square of the energy and the total angular momentum $(J)$ is linear, and is the Regge-Chew-Frautschi relationship $M_J^2=M_{J^\prime}^2+\alpha^2(J-J^\prime)$. Here, $J$ and $J'$ denote the total angular momentum of the discussed states. $M_J$ and $M_{J^\prime}$ are the masses of these states with quantum numbers $J$ and $J^\prime$, respectively. 
We need to emphasize that in this work we do not consider the possible admixture of the pseudoscalar glueball. When the possible admixture of the pseudoscalar 
glueball is substantial, the related mass formula was derived in Refs. \cite{Kataev:1981aw,Kataev:1981gr}.

\begin{figure}[htbp]
\centering
\begin{tabular}{cc}
\hspace{-20pt}
\includegraphics[width=0.45\textwidth,scale=0.35]{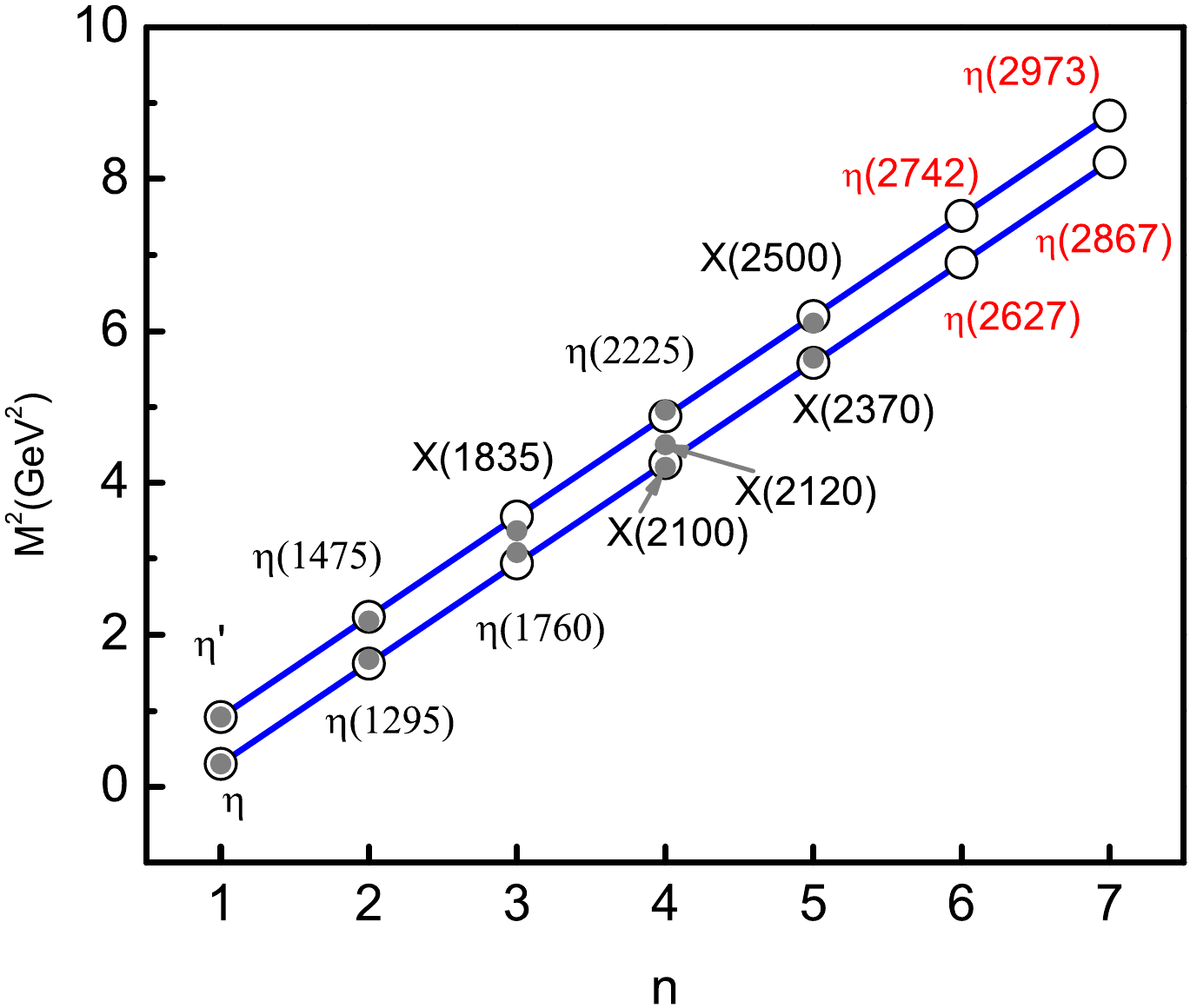}
\end{tabular}
\caption{Two Regge trajectories for the discussed $\eta/\eta^{\prime}$-like states. Here, we take $\mu^2=1.32$ GeV$^2$ suggested in Refs. \cite{Yu:2011ta,Wang:2012wa}. Here, the experimental data are from the PDG \cite{Tanabashi:2018oca}.}\label{figRegge}
\end{figure}

For discussing the fifth and the sixth radial excited states of pseudoscalar mesons, we adopt another version of Regge trajectory analysis, which was adopted in the study of different light meson systems \cite{Yu:2011ta,Wang:2012wa,Ye:2012gu,He:2013ttg,Ebert:2009ub,Ebert:2009ua,Chen:2015iqa,Wang:2019qyy}. The relation of the mass and the radial quantum number $n$ satisfies
$M^2=M_0^2+(n-1)\mu^2$, where $M_0$ is the mass of ground state and $M$ is the mass of excited state with radial quantum number $n$. $\mu^2$ represents the slope of the trajectory, and its value range is $\mu^2=1.25\pm0.15$ GeV$^2$.

In Fig. \ref{figRegge}, we plot two Regge trajectories with $\mu^2=1.32$ GeV$^2$ \cite{Yu:2011ta,Wang:2012wa}, by which we predict that $\eta(6S)$, $\eta^\prime(6S)$, $\eta(7S)$ and $\eta^\prime(7S)$
have masses of {2627}, 2742, 2867, and 2973 MeV, respectively.

In Ref. \cite{Ablikim:2016itz}, BESIII indicated that the evidence of a $\eta/\eta^\prime$-like state around 2.6 GeV may exist in the $\eta^\prime \pi^+\pi^-$ invariant mass spectrum. {This possible enhancement structure may correspond to the $\eta(2627)$ as $\eta(6S)$  predicted in this work.} In the following section, we will further discuss the decay behavior of $\eta(2627)$, which can provide valuable information to future experimental investigation.
Additionally, we will illustrate the decay properties of the remaining three predicted $\eta/\eta^\prime$ mesonic states $\eta^\prime(6S)=\eta(2742)$, $\eta(7S)=\eta(2867)$, and $\eta^\prime(7S)=\eta(2973)$.

\section{Decay properties of the four discussed $\eta/\eta^{\prime}$ excitations}\label{sec3}

For calculating the OZI-allowed decay behavior of these four predicted $\eta/\eta^{\prime}$ higher radiation excitations,
we utilize the quark pair creation (QPC) model \cite{Anisovich:2005wf,Roberts:1992js,Blundell:1996as}. The QPC model was first proposed by Micu \cite{Micu:1968mk} and then further developed by the Orsay group \cite{LeYaouanc:1972vsx,LeYaouanc:1973ldf,LeYaouanc:1974cvx,LeYaouanc:1977gm,LeYaouanc:1977fsz}. It was widely used to study the OZI-allowed strong decay of hadrons \cite{Yu:2011ta,Li:2008et,Li:2008we,Li:2008mza,Pan:2016bac,He:2013ttg,Liu:2010tr,Wang:2012wa,Guo:2019wpx,Ye:2012gu,Chen:2015iqa}.
In the QPC model, when a meson decay occurs, a quark-antiquark pair is created from vacuum with the quantum number $J^{PC}=0^{++}$, which can be combined with the corresponding antiquark and quark in the initial meson to form two final mesons.

A transition operator $T$ is introduced to describe a $q\bar{q}$ pair creation from the vacuum
\begin{eqnarray}
\label{T}
 T&=&-3\gamma\sum_{m}\langle1m;1~-m|00\rangle\int d\textbf{k}_3d\textbf{k}_4\delta^3(\textbf{k}_3+\textbf{k}_4) \nonumber \\
&&\times \mathcal{Y}_{1m}\left(\frac{\textbf{k}_3-\textbf{k}_4}{2}\right)\chi^{34}_{1,-m}
 \phi_0^{34}\omega^{34}_0 d^{\dag}_{3i}(\textbf{k}_3)b^{\dag}_{4j}(\textbf{k}_4).
\end{eqnarray}
Here, parameter $\gamma$ represents the probability
that a quark-antiquark pair is creation from the vacuum. $\textbf{k}_3$ and $\textbf{k}_4$ denote the 3-momenta of quark and antiquark created from the vacuum, respectively. $\phi_0^{34}=(u\bar{u}+d\bar{d}+s\bar{s})/\sqrt{3}$ describes the flavor singlet, and $\omega_0^{34}=\delta_{\alpha_3\alpha_4}/\sqrt{3}(\alpha=1,2,3)$ denotes the color singlet. $\chi^{34}_{1,-m}$ is a spin triplet state. $i$ and $j$ are the $SU(3)$ color indices of the created quark pairs from the vacuum. $\mathcal{Y}_{\ell m}(\mathbf{k})\equiv|\mathbf{k}|^{\ell}Y_{\ell m}(\theta_K,\phi_K)$ represents the $\ell$th solid harmonic polynomial.

By the Jacob-Wick formula \cite{Jacob:1959at,Chung:1971ri,Blundell:1996as}, the decay amplitude can be expressed as
\begin{eqnarray}\label{regge1}
\mathcal{M}_{LS}(K)&=&\frac{4\pi\sqrt{2L+1}}{2J_A+1}\sum_{M_{J_B},M_{J_C}}\langle L0SM_{J_A}|J_AM_{J_A}\rangle \nonumber \\
 &&\times\langle J_BM_{J_B}J_CM_{J_C}|SM_{J_A}\rangle
\mathcal{M}^{M_{J_A}M_{J_B}M_{J_C}}(K\hat{z}),
\end{eqnarray}
and the general decay width reads
\begin{equation}
\Gamma_i=\pi^2\frac{|\textbf{K}|}{M_A^2}\sum_{LS}|\mathcal{M}_{LS}|^2.
\end{equation}
We approximately take the sum of the partial decay widths as the total width of the discussed states in this work, $\Gamma_{\mathrm{Total}}=\sum_i \Gamma_i$.

In the concrete calculation, the harmonic oscillator wave function
$\Psi_{nlm}(R,\mathbf{p})=\mathcal{R}_{nl}(R,\mathbf{p})\mathcal{Y}_{lm}(\mathbf{p})$ is introduced, where the parameter $R$ can be determined by reproducing the realistic root-mean-square radius of the corresponding meson state.

\begin{figure}[h]
\centering
\begin{tabular}{cc}
\hspace{-20pt}
\includegraphics[width=0.5\textwidth,scale=0.35]{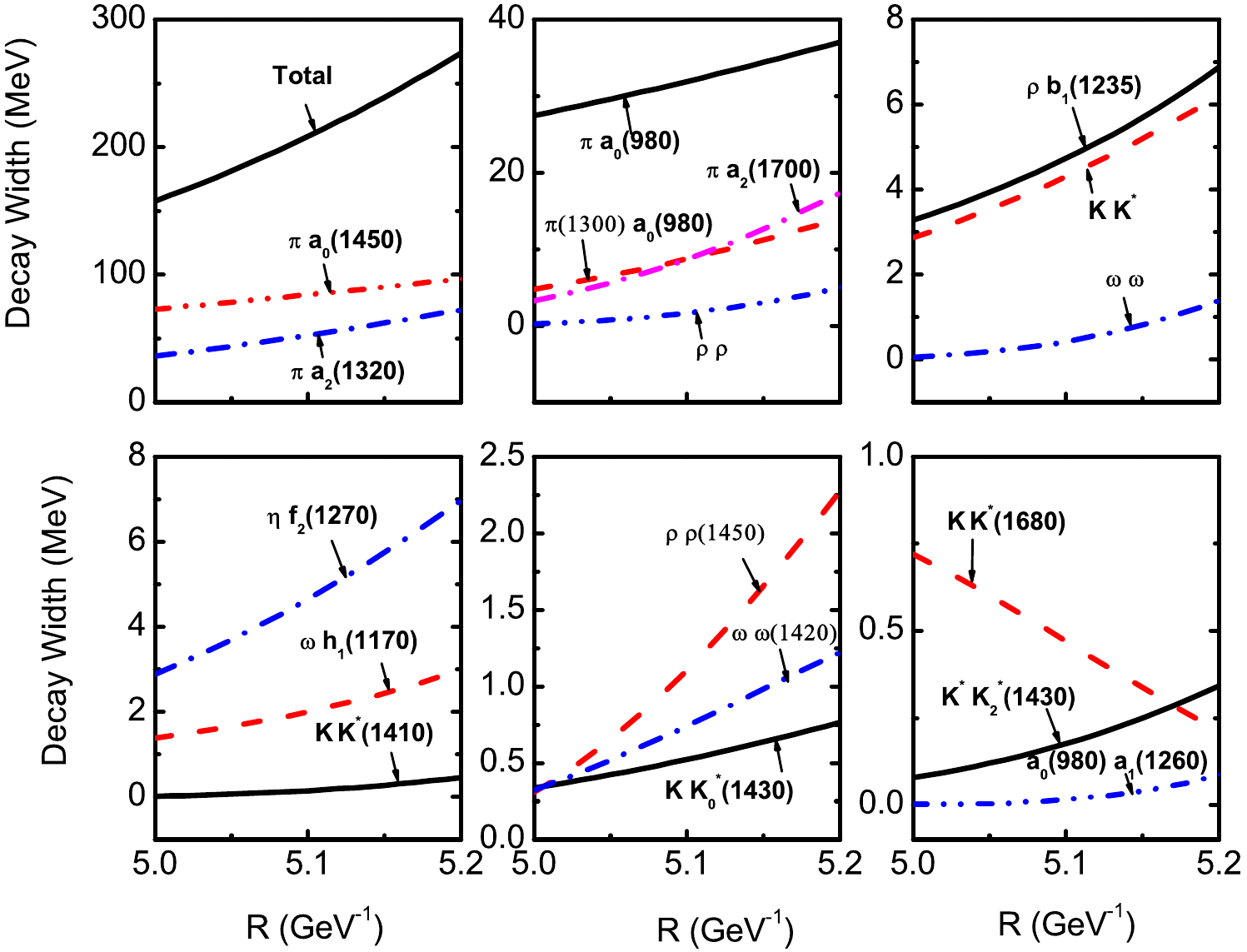}
\end{tabular}
\caption{{$R$ dependence of the total and partial decay widths of $\eta(2627)$ as $\eta(6S)$}. Some tiny channels are not drawn.}\label{2627}
\end{figure}

\begin{figure}[h]
\centering
\begin{tabular}{cc}
\hspace{-20pt}
\includegraphics[width=0.5\textwidth,scale=0.35]{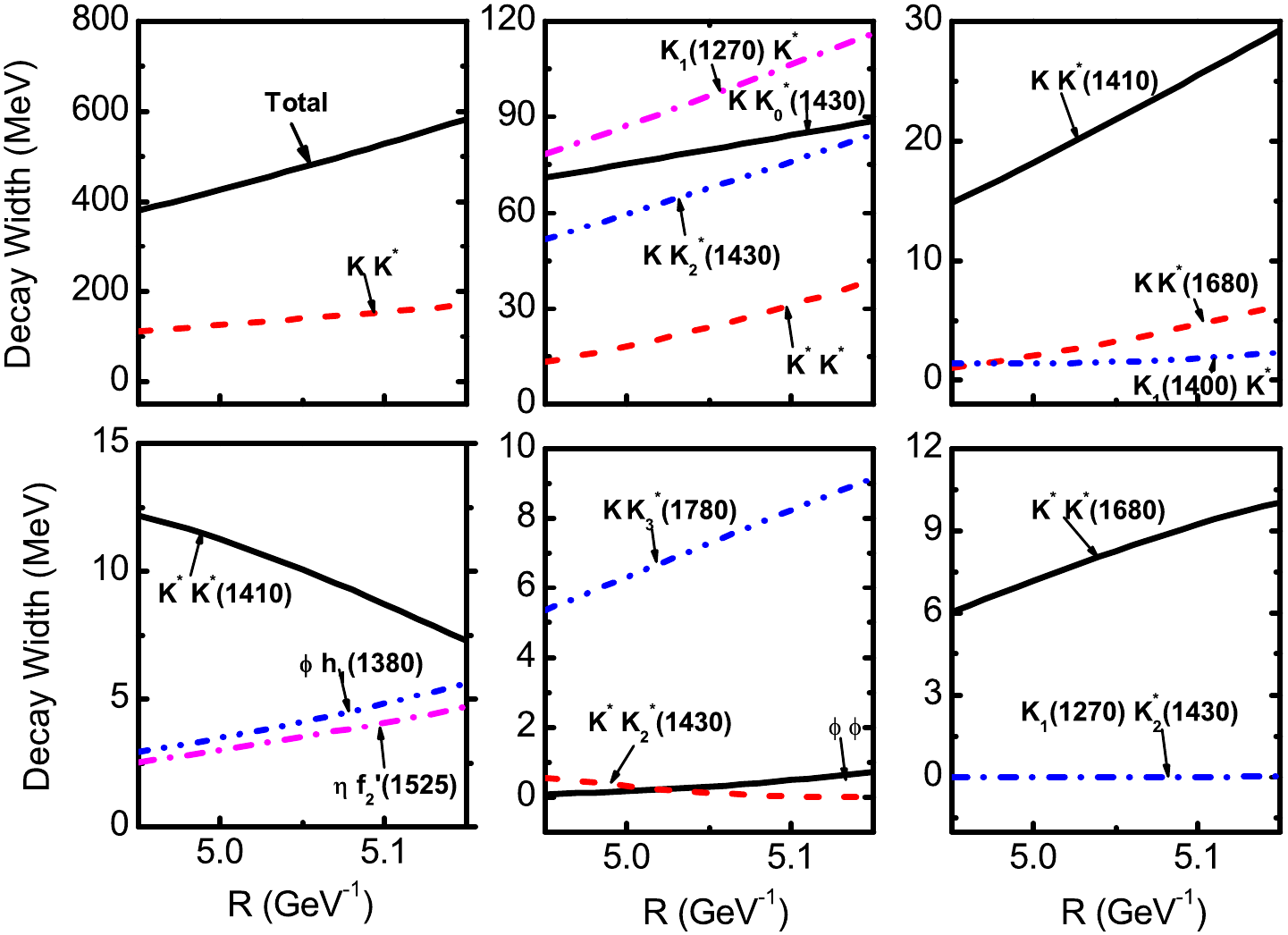}
\end{tabular}
\caption{{The $R$ dependence of total decay width and partial two-body decay widths of $\eta(2742)$ as $\eta'(6S)$}. Some tiny channels are not listed here.}\label{2742}
\end{figure}
\begin{figure}[h]
\centering
\begin{tabular}{cc}
\hspace{-20pt}
\includegraphics[width=0.5\textwidth,scale=0.35]{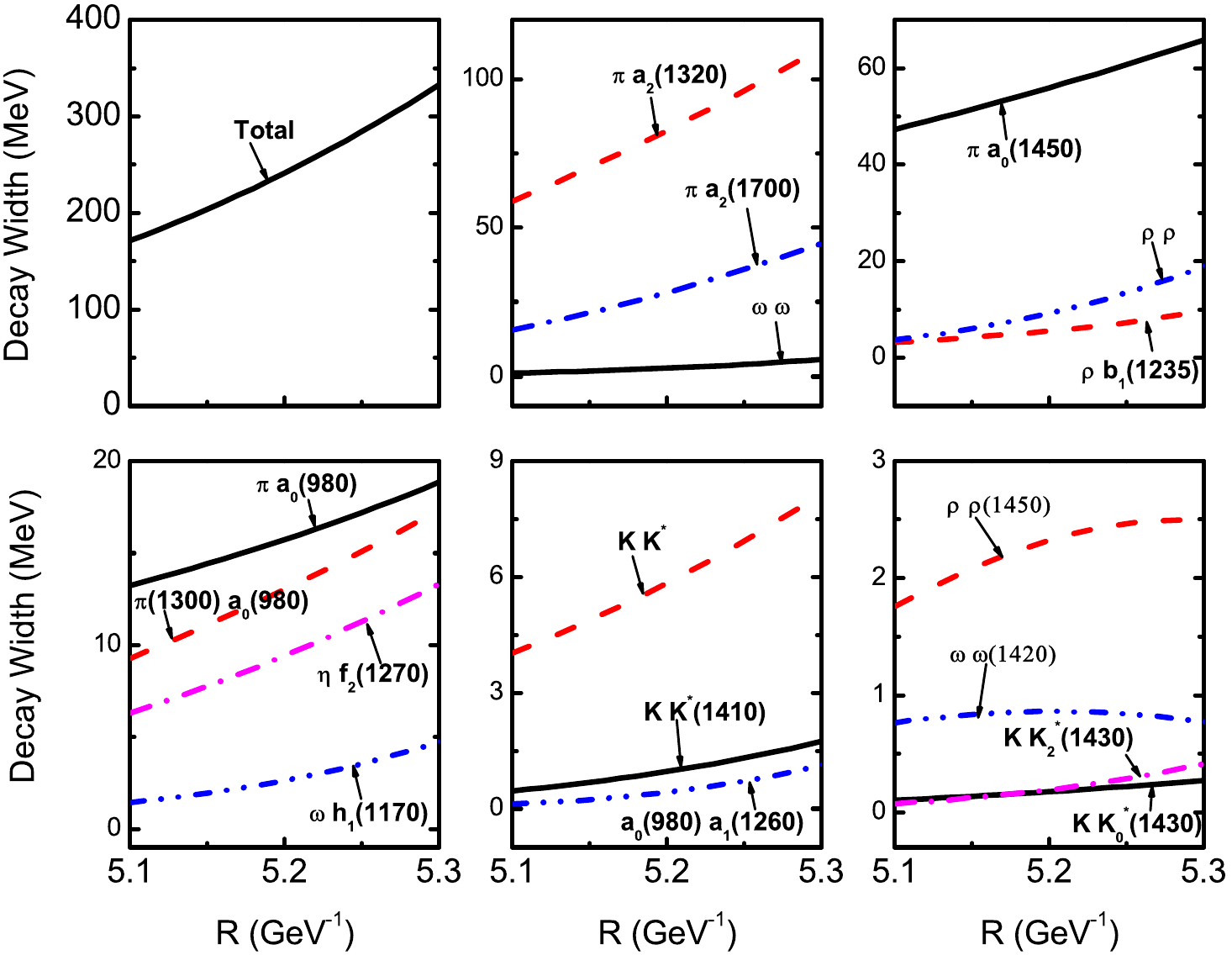}
\end{tabular}
\caption{The $R$ dependence of two-body strong decay widths of $\eta(2867)$ as a $\eta(7S)$ state. Some tiny channels are not listed here.}\label{2867}
\end{figure}

\begin{figure}[h]
\centering
\begin{tabular}{cc}
\hspace{-20pt}
\includegraphics[width=0.5\textwidth,scale=0.35]{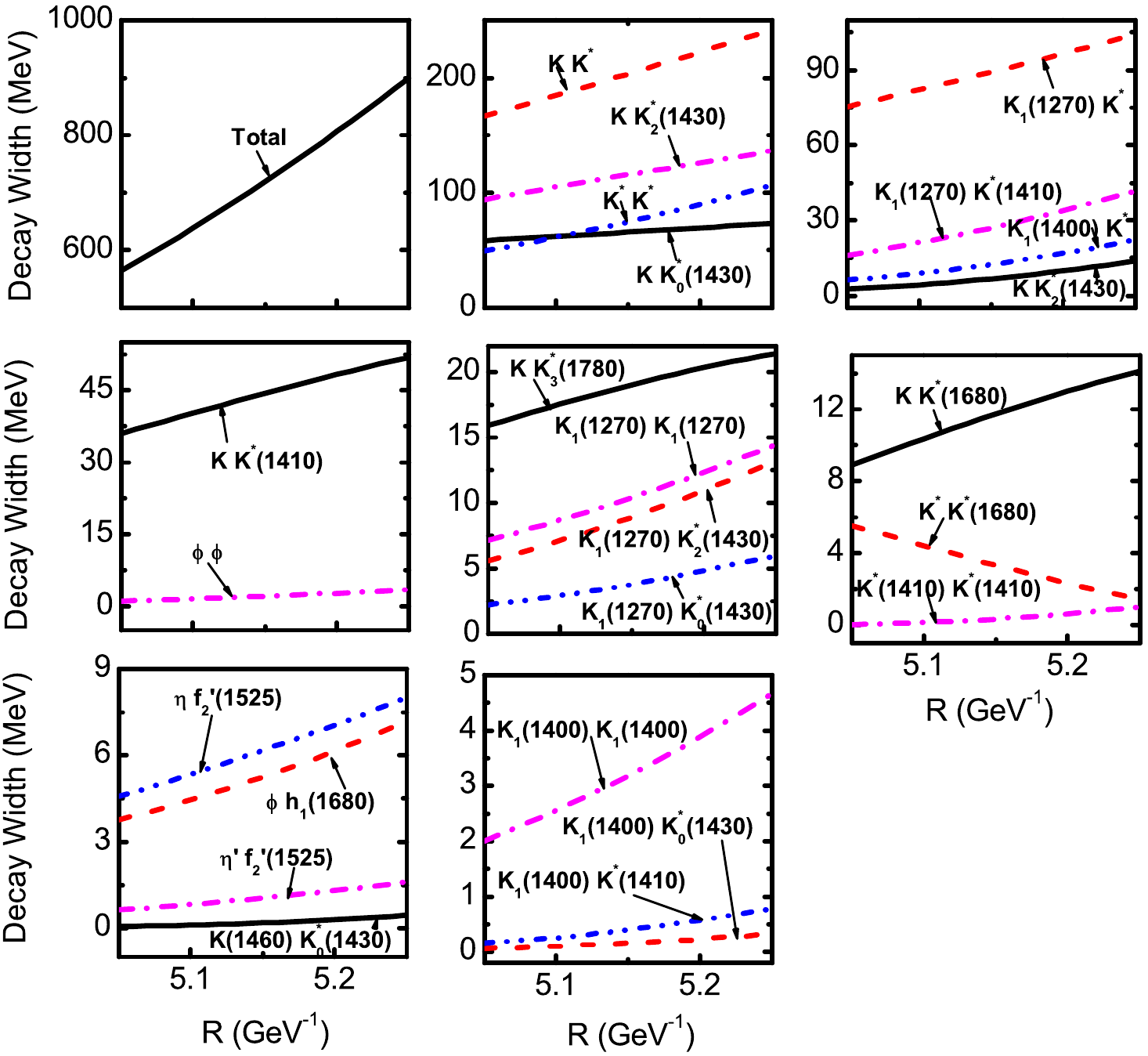}
\end{tabular}
\caption{The $R$ dependence of total decay widths and the partial two-body decay width of $\eta(2973)$ as $\eta^\prime(7S)$. Some tiny channels are not listed here.}\label{2973}
\end{figure}

In Refs. \cite{Yu:2011ta,Wang:2017iai}, the Lanzhou group performed a systematic study on pseudoscalar mesonic states by combing with these observed pseudoscalar $X$ states shown in the $\eta^\prime \pi^+\pi^-$ invariant mass spectrum \cite{Ablikim:2005um,Ablikim:2010au,Ablikim:2016itz}. These investigations enforce the possibility of categorizing these reported pseudoscalar $X$ states into a pseudoscalar meson family. Along this line, we further present the decay properties of four $\eta/\eta^{\prime}$ excitations [$\eta(2627)$, $\eta(2742)$, $\eta(2867)$ and $\eta(2973)$] listed in Fig. \ref{figRegge}. The reader may consult Figures \ref{2627}-\ref{2973} for more information of their two-body strong decay\footnote{For these discussed higher $\eta/\eta^\prime$ states, they can also decay into lower pseudoscalar ones in the same trajectory. In our calculation, we find that the partial width of this kind of decay is not sizable. For example, the decay width of $\eta(2627)\to \eta(1295)f_2(1270)$ is around 0.01 MeV. Thus, in Figures \ref{2627}-\ref{2973}, we do not list these channels.}, which is dependent on the $R$ range\footnote{In Ref. \cite{Yu:2011ta}, when $X(2370)$ is treated as $\eta(5S)$ and $\eta'(5S)$, the $R$ values corresponding to the intersection of theoretical result and experimental center value are 5.0 GeV$^{-1}$ and 4.95 GeV$^{-1}$, respectively. Usually, the $R$ value of the state in the $\eta/\eta^\prime$ meson family becomes larger with increasing the radial quantum number. Thus, for the discussed $\eta(6S)/\eta'(6S)$, we study their strong decay behaviors by taking the $R$ range (5.0-5.2) GeV$^{-1}$/(4.95-5.15) GeV$^{-1}$. And then, for the discussed $\eta(7S)$ and $\eta'(7S)$, $R$ value is set to be (5.1-5.3) GeV$^{-1}$ and (5.05-5.25) GeV$^{-1}$, respectively.}.
Additionally, we need to emphasize that a mixing scheme
\begin{eqnarray}\left(
\begin{matrix}
\eta(nS)\\
\eta^\prime(nS)
\end{matrix}\right)=\left(\begin{matrix}
\cos\theta_n&-\sin\theta_n\\
\sin\theta_n&\cos\theta_n
\end{matrix}\right)
\left(
\begin{matrix}
\eta_q(nS)\\
\eta_s(nS)
\end{matrix}\right)
\end{eqnarray}
should be introduced when discussing four $\eta^{(\prime)}$ mesons. Here, $|\eta_q(nS)=\frac{1}{\sqrt{2}}(|u\bar{u}\rangle+|d\bar{d}\rangle)$
and $|\eta_s(nS)=|s\bar{s}\rangle$ are the flavor wave functions. For the fifth and the sixth radial excitations, the information of the mixing angle is still absent.
Thus, we roughly take $4.18^{\circ}$ \cite{Wang:2017iai} in our concrete calculation under the assumption that these mixing angles for the fifth and the sixth radial excitation are same as that of the fourth radial excitation.

According to our calculation result, we summarize the main points of the total and partial decay widths of the four discussed states:

\begin{itemize}

\item As the $\eta(6S)$ state, $\eta(2627)$ has the main decay channels $\pi a_0(1450)$, $\pi a_2(1320)$, and $\pi a_0(980)$ (see Fig.\ref{2627}). Since $a_0(1450)$ may decay into $\eta\pi$ and $\pi\eta^\prime(958)$ with the branching ratios $9.3\%$ and $3.3\%$, respectively, three-body decays $\eta(2627)\to \eta\pi^+\pi^-$ and $\eta(2627)\to \eta^\prime\pi^+\pi^-$ should be sizable. Thus, searching for $\eta(2627)$ via the $\eta\pi^+\pi^-$ and $\eta^\prime\pi^+\pi^-$ final states is suggested. We notice BESIII's measurement of $J/\psi\to\gamma\eta^\prime\pi^+\pi^-$, where the event accumulation around 2624 MeV exists in the $\eta^\prime\pi^+\pi^-$ invariant mass spectrum \cite{Ablikim:2016itz}. This enhancement event is consistent with the predicted $\eta(2627)$. Besides giving the partial decay widths of $\eta(2627)$, we also obtain the total width of $\eta(2627)$, which is in the range of $157.8-274.0$ MeV. With 10 billion $J/\psi$ events collected in 2019 from BESIII, we have reason to believe that the predicted $\eta(2627)$ as the fifth radial excitation of the $\eta/\eta^\prime$ meson family can be established in experiment, which is an important step when constructing the $\eta/\eta^\prime$ meson family.

\item
In the following, we discuss the partner of $\eta(2627)$, which corresponds to $\eta(2742)$ as $\eta^\prime(6S)$. Our result shows that $\eta(2742)$ is a broad state with width $396.3-590.1$ MeV, which may result in the difficulty to identifying it in the experiment. Its dominant decay mode includes $KK^*$, while $K_1(1270)K^*$, $KK_0(1430)$, $KK_2^*(1430)$ and $K^*K^*$ have the main contribution to the total decay width of $\eta(2742)$. In Fig. \ref{2742}, more information of its partial decay widths can be found.

\item
From the analysis of Regge trajectory (see Fig.\ref{figRegge}), $\eta(7S)$ has mass 2867, which is referred to as $\eta(2867)$ in this work. In Fig. \ref{2867}, we present its total and partial decay widths.
Similar to $\eta(2625)$ mentioned above, $\eta(2867)$ mainly decays into $\pi a_2(1320)$ and $\pi a_0(1450)$. But the width of $\eta(2867)$ is lightly broader than that of $X(2625)$.
Here, the suggested ideal channel of searching for $\eta(2867)$ is still $\eta^\prime\pi^+\pi^-$, which can be from $J/\psi\to \gamma \eta^\prime\pi^+\pi^-$. It is obvious that BESIII should try to hunt it with more experimental data.

\item The
$\eta(2973)$ as a $\eta^\prime(7S)$ state should have the decay behavior shown in Fig. \ref{2973}, where its dominant decay channel is $KK^*$. The total decay width of $\eta(2973)$ is very broad. Thus, it is difficult to discover such broad structure in the experiment. In addition, we have to face the fact that $\eta(2973)$ almost overlaps with $\eta_c(1S)$.

\end{itemize}

\section{Production relevant to the discussed $\eta^{(\prime)}(nS)$ induced by a pion or kaon }\label{sec4}

Until now, the $\eta^{(\prime)}$ mesons mostly were observed through the $J/\psi$ radiative decay process. Searching for them in different reaction platforms is an interesting issue, which may provide more abundant information to decode these states.
It is well known that the pion-proton and kaon-proton scattering processes are effective experimental tools in exploring light hadrons. A typical example is $\eta(1295)$, which was first observed in the pion-proton scattering process $\pi^- p\to n\eta\pi^+\pi^-$ \cite{Stanton:1979ya}.
Therefore, pion-proton and kaon-proton scattering could be a peculiar way to investigate the $\eta^{(\prime)}$ mesons.
Based on this motivation, in this work, we will explore the productions of $\eta^{(\prime)}(6S)$ and $\eta^{(\prime)}(7S)$ via the pion-proton and kaon-proton scattering processes, where the effective Lagrangian approach is adopted. These calculated results are valuable for the further relevant experimental exploration, where several concrete experiments include J-PARC ~\cite{Nagae:2008zz,Kumano:2015gna}, COMPASS\cite{Nerling:2012er}, and SPS@CERN \cite{Velghe:2016jjw}.

\begin{figure}[htbp]
  \centering
  \begin{tabular}{cc}
  \includegraphics[width=115pt]{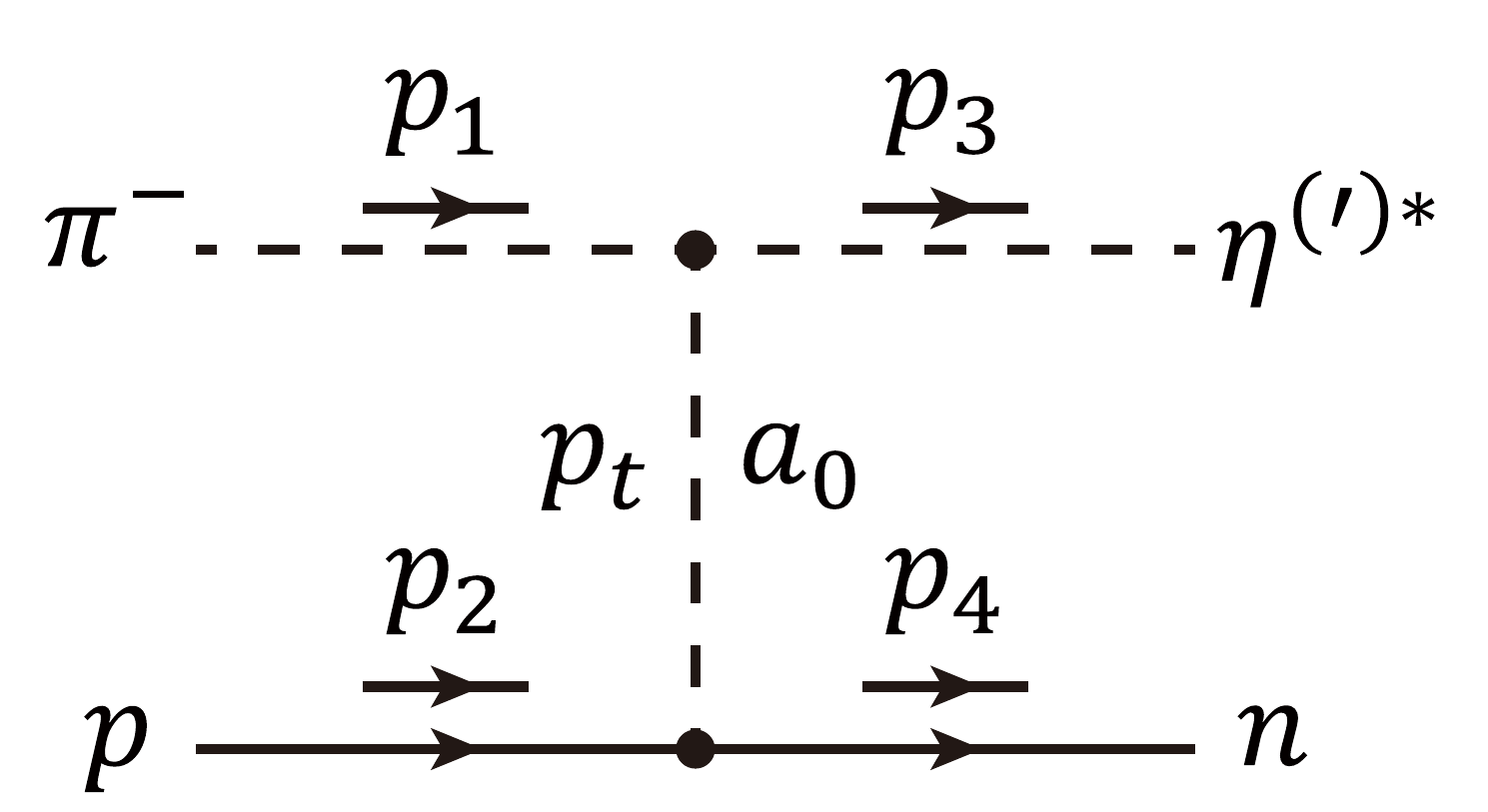}&\includegraphics[width=115pt]{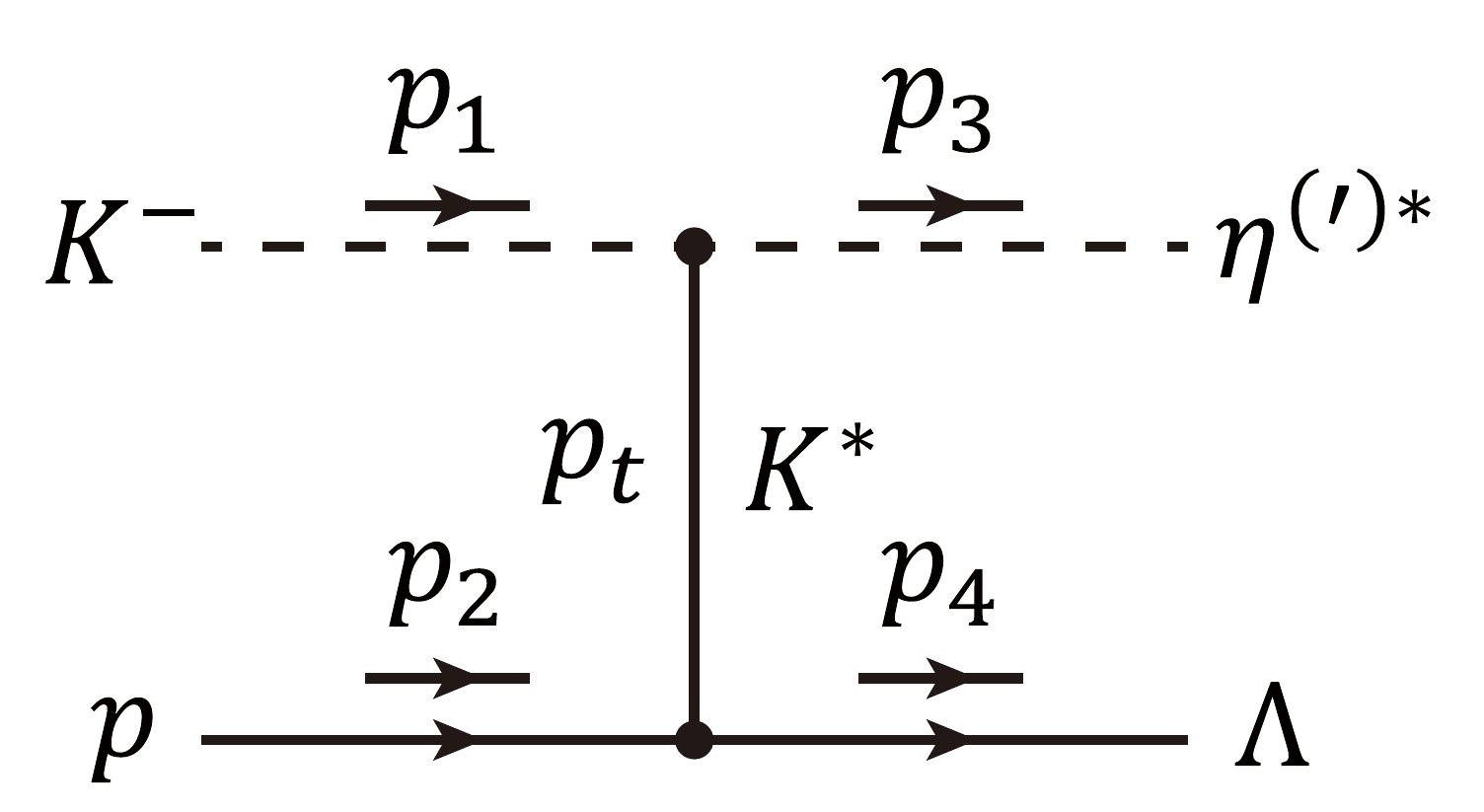}\\
  \end{tabular}
  \caption{Feynman diagrams for the $\pi^{-}p\rightarrow \eta^{(\prime)*} n$ reaction (left) and the $K^- p\rightarrow \eta^{(\prime)*} \Lambda$ reaction (right), where $\eta^{(\prime)*}$ donate $\eta(nS)$ or $\eta^{\prime}(nS)$ with $n=6$, 7.}\label{Fey-dia}
\end{figure}

The Feynman diagram of these discussed higher radial excitations of the pseudoscalar meson family produced via pion and kaon induced reactions on a proton target is illustrated in Fig. \ref{Fey-dia}, where we only consider the $t$ channel diagram.
Besides, the contributions from $s$ channel and $u$ channel are not considered in this work, since the $s$ channel is usually negligibly small and the $u$ channel always concentrates at backward angles.

For the  $\pi^- p\rightarrow \eta^{(\prime)*} n$ reaction, we take the relevant effective Lagrangians \cite{Gokalp:2002xi,Kirchbach:1995ep}
\begin{eqnarray}
\mathcal{L}_{a_0\pi\eta^{(\prime)*}}&=&g_{a_0\pi\eta(nS)}\vec{a_0} \cdot \vec{\pi} \,\eta^{(\prime)*},\\
\mathcal{L}_{a_0 N N}&=&g_{a_0 NN}\bar{N} \vec{a_0} \cdot \vec{\tau} N,
\end{eqnarray}
where $a_{0}$, $\eta^{(\prime)*}$, $\pi$, and $N$ donate the $a_{0}(980)$, $\eta(nS)$, or $\eta^{\prime}(nS)$ with $n=6$, 7, pion, and nucleon fields, respectively.
The coupling constants $g_{a_0\pi\eta^{(\prime)*}}$ ($g_{a_0\pi\eta(6S)}=1.28$ GeV, $g_{a_0\pi\eta(7S)}=0.93$ GeV, $g_{a_0\pi\eta^{\prime}(6S)}=0.09$ GeV, and $g_{a_0\pi\eta^{\prime}(7S)}=0.06$ GeV) can be determined by the decay width of $\eta^{(\prime)*}$ decaying into $\pi a_{0}(980)$, which is calculated by the QPC model in Sec. \ref{sec3}. Besides, for the coupling constant $g_{a_0 NN}$, we adopt $g_{a_0 NN}^2 /4 \pi=1.075$, which is implied by the Bonn one-boson exchange model for the nucleon-nucleon interaction \cite{Machleidt:1987hj}.

For the $K^- p\rightarrow \eta^{(\prime)*}\Lambda$ reaction, the involved effective Lagrangians include \cite{Oh:2006hm,Oh:2006in,Wang:2019uwk}
\begin{eqnarray}
\mathcal{L}_{K^* K\eta^{(\prime)*}}&=&ig_{K^* K \eta^{(\prime)*}}(K\partial^\mu \eta^{(\prime)*}-\eta^{(\prime)*} \partial^{\mu}K)K^*_\mu,\\ \nonumber
\mathcal{L}_{K^* N \Lambda}&=&-g_{K^* N \Lambda}\bar{\Lambda}\left(\gamma_\mu-\frac{\kappa_{K^* N \Lambda}}{2m_{N}}\sigma_{\mu\nu}\partial^\nu\right)K^{*\mu} N+\textrm{H.c.},\\
\end{eqnarray}
where the coupling constants $g_{K^* N \Lambda}=-4.26$ and $\kappa_{K^* N \Lambda}=2.66$ are adopted in Ref. \cite{Stoks:1999bz}, which are from the Nijmegen potential. Additionally, the coupling constants $g_{K^* K \eta^{(\prime)*}}$ ($g_{K^* K \eta(6S)}=0.02$, $g_{K^* K \eta(7S)}=0.02$, $g_{K^* K \eta^{\prime}(6S)}=0.11$, and $g_{K^* K \eta^{\prime}(7S)}=0.11$) are estimated by the decay width of $\eta^{(\prime)*} \rightarrow K K^*$, which is also obtained by the QPC model (see Sec. \ref{sec3}).

With the above preparation, the amplitudes of these discussed reactions shown in Fig. \ref{Fey-dia} can be written as
\begin{eqnarray}\nonumber
i\mathcal{M}_{\pi^- p\rightarrow n\eta^{(\prime)*}}&=&\sqrt{2}g_{a_0 \pi \eta^{(\prime)*}}g_{a_0 N N}\bar{u}(p_4)u(p_2)\frac{i}{t-m_{a_0}^2}F_{a_0}^2(t),\\ \label{AmplitudePionn}
\\ \nonumber
i\mathcal{M}_{K^{-}p\rightarrow\Lambda\eta^{(\prime)*}}&=&-ig_{K^* K \eta^{(\prime)*}}g_{K^* N \Lambda}\bar{u}(p_4)\left[\gamma_{\mu}-\frac{\kappa_{K^* N \Lambda}}{2m_{N}}\sigma_{\mu\nu}(i p_{t}^{\nu})\right]\\ \nonumber
&&\times u(p_2)\frac{i\tilde{g}^{\mu\alpha}}{p_{t}^{2}-m_{K^*}^2}\left[(ip_{3}^{\alpha})-(-ip_{1}^{\alpha})\right]F_{K^*}^2(t),\\ \label{AmplitudeKaonn}
\end{eqnarray}
where $t=p_t^2=(p_3-p_1)^2$, $\tilde{g}^{\mu\alpha}=-g^{\mu\alpha}+p_{t}^{\mu} p_{t}^{\alpha}/m_{K^{*}}^2$, and $F_{x}(t)$ is the form factor of $t$-channel exchange for each interaction vertex, which is taken as the monopole form ~\cite{Machleidt:1987hj,Machleidt:1989tm}
\begin{eqnarray}
F_{x}(t)=(\Lambda_{x}^2-m_{x}^2)/(\Lambda_{x}^2-t)
\end{eqnarray}
 in this work.
The cutoff $\Lambda_{x}$ in the form factor can be parametrized as $\Lambda_{x}=m_{x}+\alpha \Lambda_{QCD}$ with $\Lambda_{QCD}=220$ MeV. In general, the value of parameter $\alpha$ is taken around 1~\cite{Cheng:2004ru,Wang:2019qyy}. Thus, we also take $\alpha=1$ in our calculation.

Here, we also introduce the Reggeized treatment to the $t$ channel in order to better describe the behavior of the hadron production at high momentum ~\cite{Wang:2019qyy,Collins:1977,Guidal:1997hy,Galata:2011bi,Wang:2017qcw}.
To the Reggeized treatment for the $t$-channel meson exchange, we only need to replace the form factor in the Feynman amplitudes in Eqs. (\ref{AmplitudePionn}) and (\ref{AmplitudeKaonn}) as

\begin{eqnarray}\nonumber
F_{a_0}^2(t)&\rightarrow&\left(\frac{s}{s_{\textrm{scale}}}\right)^{\alpha_{a_{0}}(t)}
\frac{\pi\alpha_{a_{0}}^{\prime}(t)}{\Gamma[1+\alpha_{a_{0}}(t)]\textrm{sin}[\pi\alpha_{a_{0}}(t)]}\\ \label{Ragge1}
&&\times\left(1+\xi e^{-i\pi\alpha_{a_{0}}(t)}\right)\left(t-m_{a_{0}}^{2}\right),\\ \nonumber
F_{K^*}^2(t)&\rightarrow&\left(\frac{s}{s_{\textrm{scale}}}\right)^{\alpha_{K^*}(t)-1}
\frac{\pi\alpha_{K^*}^{\prime}(t)}{\Gamma[\alpha_{K^{*}}(t)]\textrm{sin}[\pi\alpha_{K^{*}}(t)]}\\ \label{Ragge2}
&&\times\left(1+\xi e^{-i\pi\alpha_{K^{*}}(t)}\right)\left(t-m_{K^{*}}^{2}\right).
\end{eqnarray}
The scale factor $s_{\textrm{scale}}$ is fixed at 1 GeV, and we set the signature $\xi=1$ for the $a_0$ exchange and $\xi=-1$ for the $K^{*}$ exchange.
The Regge trajectories of $\alpha_{a_{0}}(t)$ and $\alpha_{K^{*}}(t)$ read as
\begin{eqnarray}
\alpha_{a_{0}}(t)&=&-0.5+0.6t,\\
\alpha_{K^{*}}(t)&=&1+0.85\left(t-m_{K^*}^2\right),
\end{eqnarray}
respectively.
Besides, the Gamma function can suppress poles of the $\sin[\pi \alpha(t)]$ when the $\alpha(t)\leq 0$, and the poles of $\alpha(t)> 0$  can be avoided automatically since $t<0$ for the $t$ channel leads to $\alpha(t)<1$.

Now, all parameters have been determined. Then, we can calculate the cross sections of the productions of these four pion and kaon induced reactions on a proton target.
For the $2\rightarrow 2$ reaction process, the differential scattering cross section can be expressed as
\begin{eqnarray}
\frac{d\sigma}{dt}=\frac{1}{64\pi s}\frac{1}{p_{1cm}}\overline{|\mathcal{M}|}^2,
\end{eqnarray}
where $s=(p_1+p_2)^2$ is the square of center of mass energy, $p_{1cm}$ denotes the momentum of incident pion or kaon in the center of mass frame, and the overline on $\overline{|\mathcal{M}|}^2$ indicates the square of amplitudes average over the polarizations in the initial states and the sum over the polarization in the final states.

\begin{figure}[htbp]
\centering
\begin{tabular}{cc}
\hspace{-20pt}
\includegraphics[width=0.45\textwidth,scale=0.35]{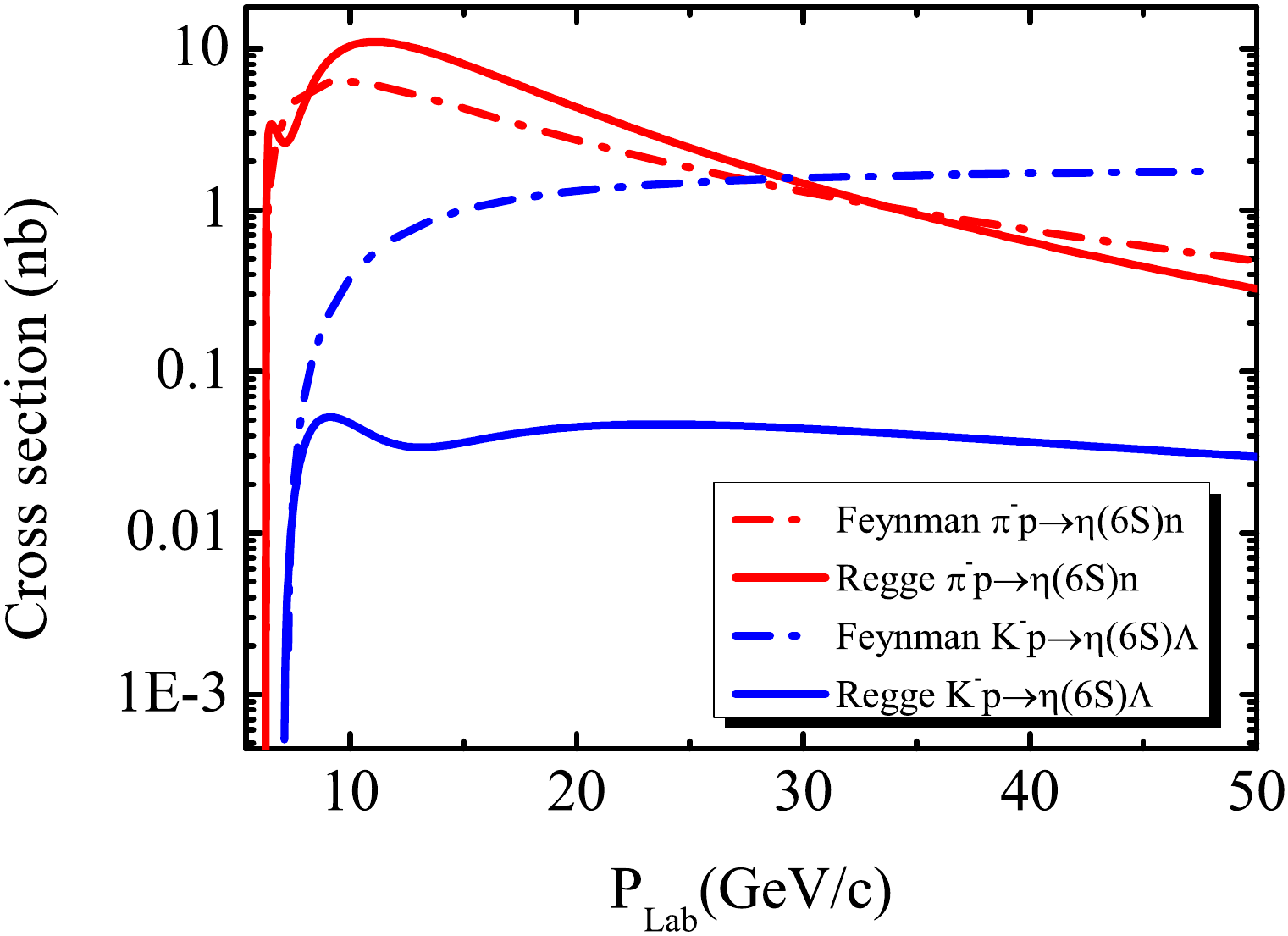}
\end{tabular}
\caption{Total cross section for the $\pi^-p\to\eta(6S)n$ and $K^-p\to\eta(6S)\Lambda$ in  the Feynman model and the Regge model.}\label{2627s}
\end{figure}

\begin{figure}[htbp]
\centering
\begin{tabular}{cc}
\hspace{-20pt}
\includegraphics[width=0.45\textwidth,scale=0.35]{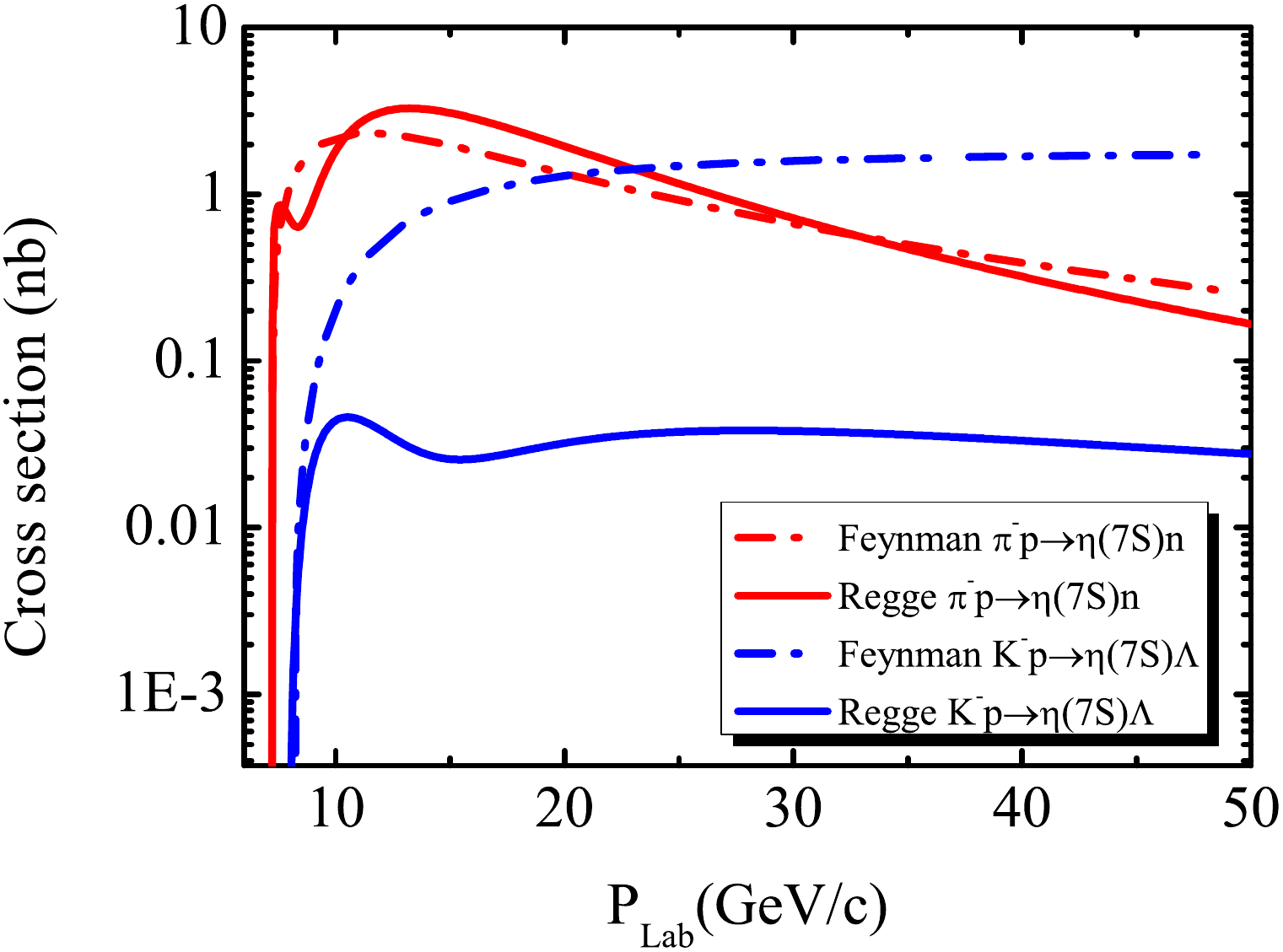}
\end{tabular}
\caption{Total cross section for the $\pi^-p\to\eta(7S)n$ and $K^-p\to\eta(7S)\Lambda$ in the Feynman model and the Regge model.}\label{2867s}
\end{figure}

In Fig. \ref{2627s}, we show the numerical result of the cross sections of $\pi^- p\rightarrow \eta(6S)n$ and $K^- p\rightarrow \eta(6S)\Lambda$ as a function of the pion and kaon momenta in the laboratory system ($P_{Lab}$), respectively.
For the $\pi^- p \rightarrow \eta(6S)n$ reaction, both of the line shapes of the total cross sections in the
Feynman model (red dotted line) and the Regge model (red solid line) sharply increase near the threshold, and then they begin to slowly decrease with increasing $P_{Lab}$.
In the Feynman model, the total cross section reaches up to a maximum of $6.3$ $nb$ at a momentum $P_{Lab}=9.5$ $\text{GeV}/c$.
But in the Regge model, the maximum of the total cross section is 11.0 $nb$ at $P_{Lab}=11.1$ $\text{GeV}/c$.
Different from the Feynman case, the line shape of the total cross section in the Regge model
decreases more rapidly with increasing $P_{Lab}$ when the total cross section has reached a maximum.

For the $K^{-}p \rightarrow \eta(6S)\Lambda$ reaction, the obtained line shapes of the total cross sections in the Feynman model (blue dotted line) and the Regge model (blue solid line) also sharply increase near the threshold, but then slowly trend to a stable value.
The line shape in the Feynman model (blue dotted line) is increasing slowly with increasing $P_{Lab}$, while the line shape in the Regge model (blue solid line) is decreasing slowly with increasing $p_{Lab}$ when the total cross section reaches up to a maximum $0.05$ $nb$ at $P_{Lab}=25.5$ $\text{GeV}/c$.
Compared to the Feynman model, the Regge model gives a smaller cross section for the $K^{-}p\rightarrow \eta(6S)\Lambda$ reaction.
Hence, the line shape difference between these two models can be applied to distinguish the role of the
Regge model in further experimentation.
Beside, we find that the total cross section of $\eta(6S)$ given by the $\pi^{-}p\rightarrow \eta(6S)n$ reaction is significantly larger than the result obtained by the $K^{-}p\rightarrow \eta(6S)\Lambda$ reaction, since $\eta(6S)$ coupling with $\pi a_{0}$ is stronger than $\eta(6S)$ interacting with $K K^{*}$.
According to our results, the pion-proton scattering may be more better platform than the kaon-proton scattering to explore the $\eta(6S)$ state.
We also suggest that the $P_{Lab}$ range with $9.0\sim12$ $\text{GeV}/c$ is a good momentum window for future experiments to hunt $\eta(6S)$ via the pion-proton scattering platform.
If experimentalist want to find $\eta(6S)$ in the kaon-proton scattering process, the $P_{Lab}$ around 25 $\text{GeV}/c$ is a suitable momentum window.

Additionally, in Fig. \ref{2867s}, we also show the numerical results of the cross sections of $\eta(7S)$ produced through the pion-proton scattering and the kaon-proton scattering, where the behavior of the lines shape of the total cross sections as a function of $P_{Lab}$ is similar to the case of $\eta(6S)$.
The total cross section of $\eta(7S)$ produced through the $\pi^{-}p\rightarrow \eta(7S) n$ reaction has a maximum 2.3 $nb$ at $P_{Lab}=11.3$ $\text{GeV}/c$ in the Feynman model and 3.3 $nb$ at $P_{Lab}=13.2$ $\text{GeV}/c$ in the Regge model.
And, in the $\pi^{-}p\rightarrow \eta(7S) n$ reaction, the total cross section tends to a stable value 1.5 $nb$ for the Feynman model and 0.04 $nb$ for the Regge model at $P_{Lab}=25$ $\text{GeV}/c$.
Hence, the $P_{Lab}$ range of $11.3\sim13.2$ $\text{GeV}/c$ and  $P_{Lab}=25$ $\text{GeV}/c$ may be a good momentum window to search for $\eta(7S)$ on the pion-proton and  kaon-proton scattering, respectively.

\begin{figure}[htbp]
\centering
\begin{tabular}{cc}
\hspace{-20pt}
\includegraphics[width=0.45\textwidth,scale=0.35]{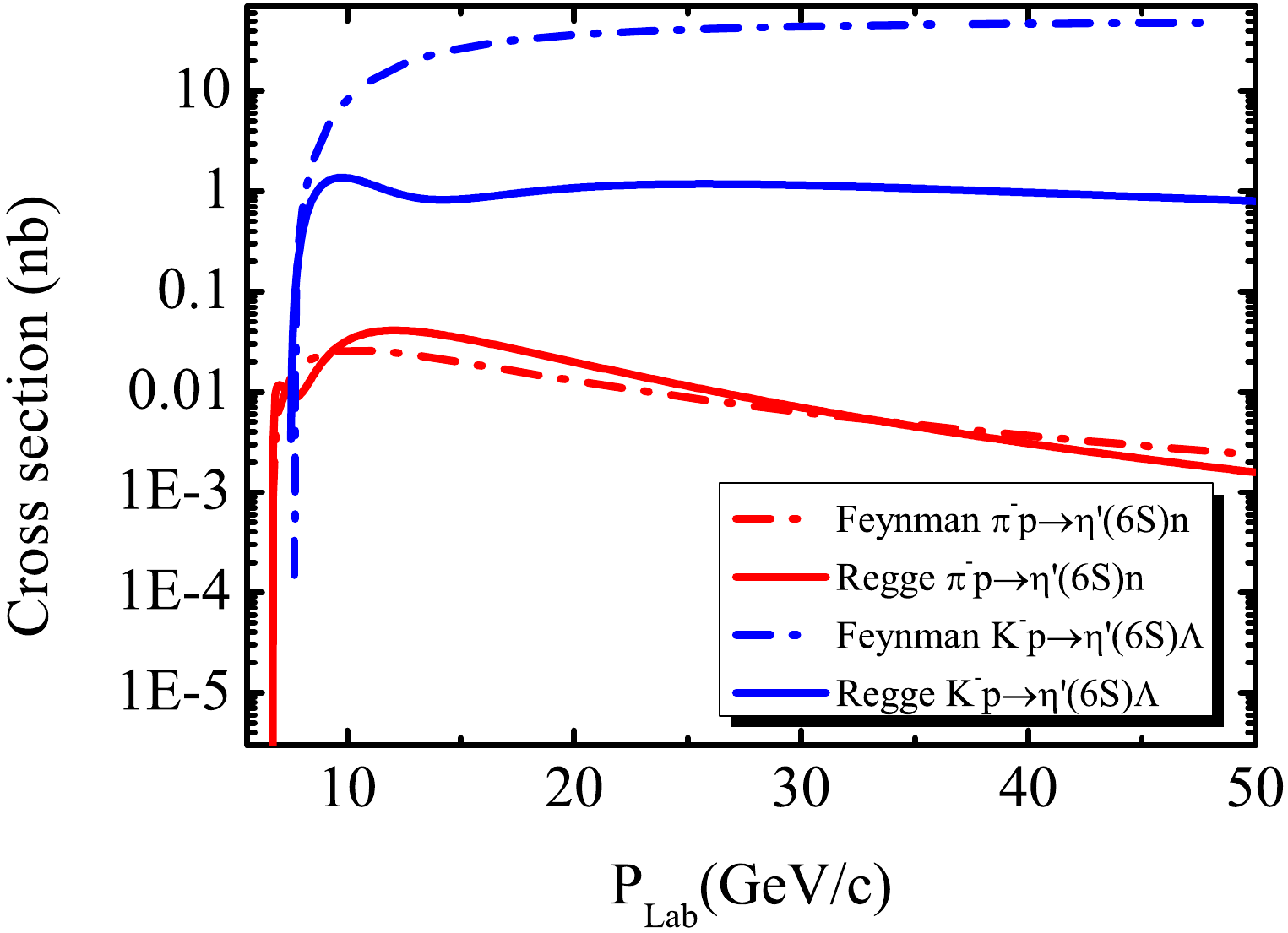}
\end{tabular}
\caption{Total cross section for the $\pi^-p\to\eta'(6S)n$ and $K^-p\to\eta'(6S)\Lambda$ in the Feynman model and Regge model.}\label{2742s}
\end{figure}

\begin{figure}[htbp]
\centering
\begin{tabular}{cc}
\hspace{-20pt}
\includegraphics[width=0.45\textwidth,scale=0.35]{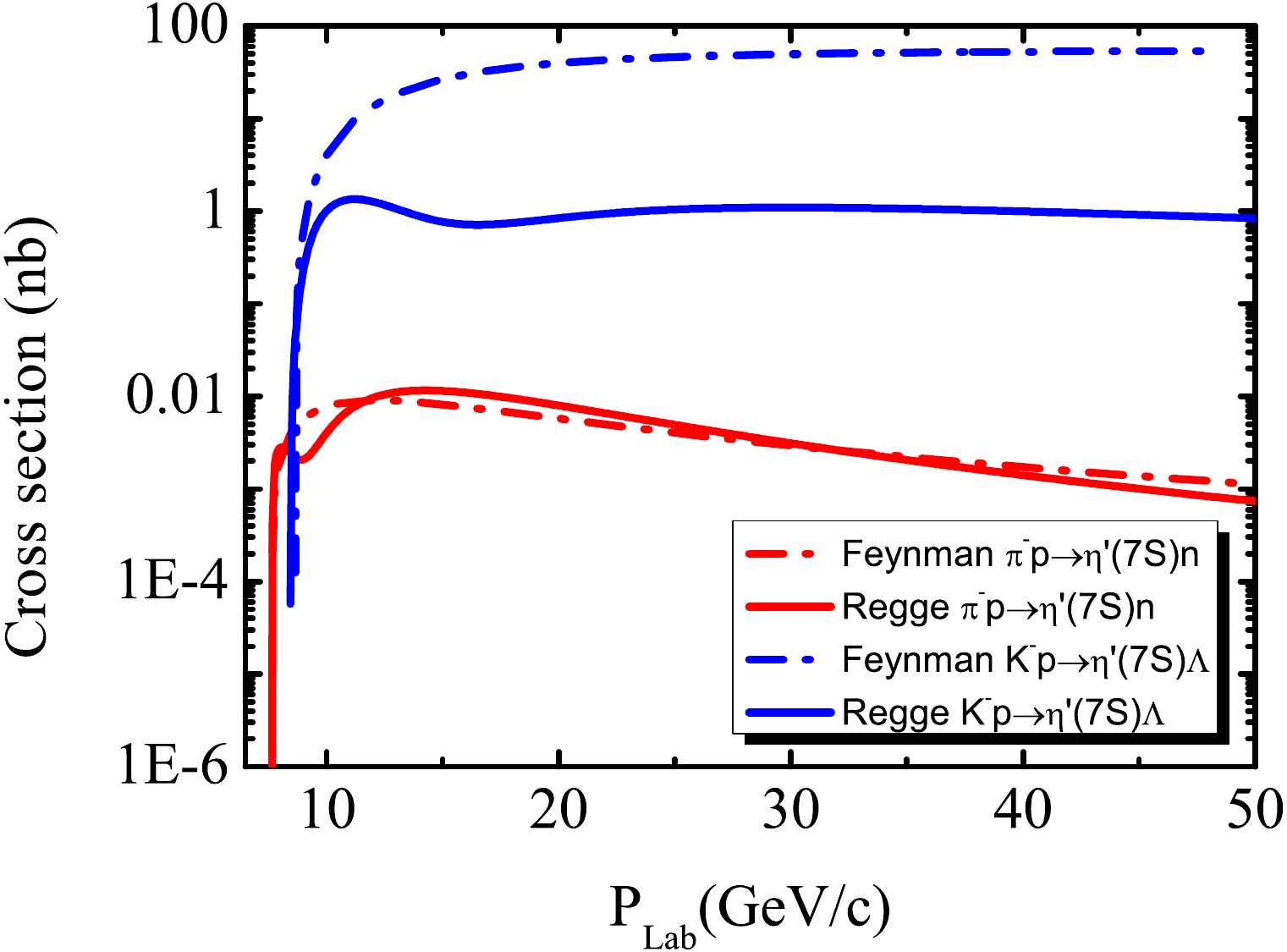}
\end{tabular}
\caption{Total cross section for the $\pi^-p\to\eta'(7S)n$ and $K^-p\to\eta'(7S)\Lambda$ in the Feynman model and Regge model.}\label{2973s}
\end{figure}

In Figs. \ref{2742s} and \ref{2973s}, we present the line shape of total production cross sections of $\eta^{\prime}(6S)$ and $\eta^{\prime}(7S)$ with the pion-proton scattering and the kaon-proton scattering, respectively.
In fact, we notice that the behavior of these line shape is similar to that of the reactions $\pi^-p\rightarrow \eta(6S)n$ and $K^- p \rightarrow \eta(7S)\Lambda$, which
is due to the similarity of the corresponding reaction amplitudes.
However, there also exists a difference, i.e., the $\eta^{\prime}(6S)$ and $\eta^{\prime}(7S)$ production cross sections obtained by the pion-proton scattering are smaller than the results obtained by the kaon-proton scattering. The main reason is that
the coupling of $\eta^{\prime}(6S)/\eta^{\prime}(7S)$ to $KK^*$ is stronger than that of them to $\pi a_{0}$.
For the $\eta^{\prime}(6S)$ ($\eta^{\prime}(7S)$) production through the pion-proton scattering, the total cross section has maximum 0.03 $nb$ (0.01 $nb$) at $P_{Lab}=10.5$ $\text{GeV}/c$ ($P_{Lab}=12.2$ $\text{GeV}/c$) in the Feynman model, (the Regge model), where the maximum is 0.04 $nb$ (0.01 $nb$) at $P_{Lab}=13.6$ $\text{GeV}/c$ ($P_{Lab}=13.6$ $\text{GeV}/c$).
The total cross sections of $\eta^{\prime}(6S)$ and $\eta^{\prime}(7S)$ in the kaon-proton scattering are much larger than the results of pion-proton scattering.
The line shape of the total cross section of the $K^-p \rightarrow \eta^{\prime}(6S)$ [$K^-p \rightarrow \eta^{\prime}(7S)$] reaction in the Feynman model sharply increases near the threshold, then slowly trends to a stable value 40 $nb$ (47 $nb$) at $P_{Lab}=25$ $\text{GeV}/c$ ($P_{Lab}=24$ $\text{GeV}/c$).
But in the Regge model, the total cross section of $\eta^{\prime}(6S)$ [($\eta^{\prime}(7S)$] in the kaon-proton scattering has a maximum of 1.4 $nb$ (1.3 $nb$) at $P_{Lab}=9.8$ $\text{GeV}/c$ ($P_{Lab}=11.2$ $\text{GeV}/c$).
Hence, the $P_{Lab}$ range of $9.8 \sim 25$ $\text{GeV}/c$ ($11.2 \sim 25$ $\text{GeV}/c$) may be a good momentum window for future experiments to research $\eta^{\prime}(6S)$ [$\eta^{\prime}(7S)$] on the kaon-proton scattering platform.  And, in the pion-proton scattering process, a suitable window is around $P_{Lab}=13.6$ $\text{GeV}/c$.

\begin{figure}[htbp]
  \centering
  \includegraphics[width=220pt]{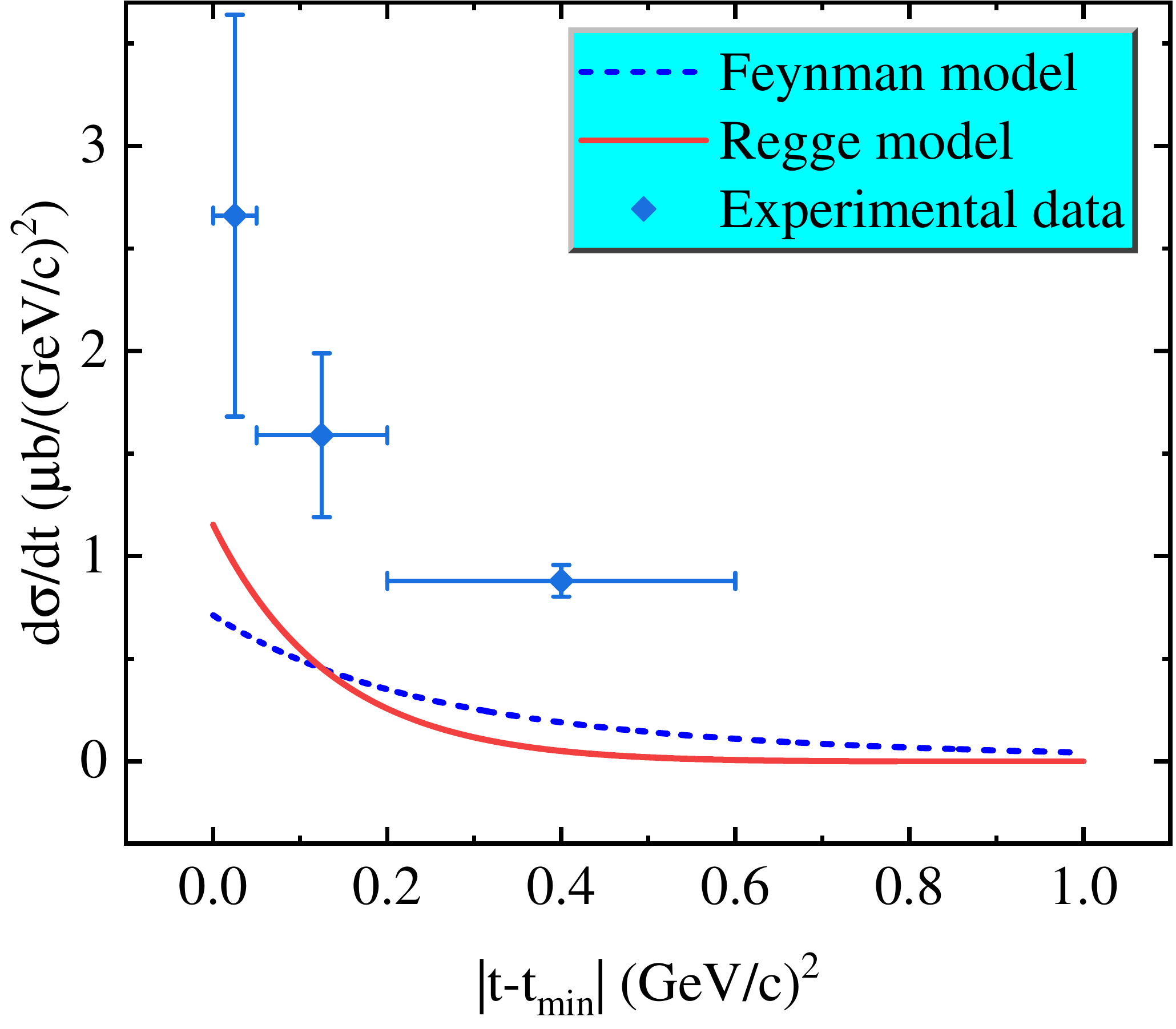}
  \caption{Differential cross section $d\sigma/dt$ as a function of $|t-t_{min}|$ for the $\pi^-p\rightarrow\eta(1295)n$ scattering at $p_{Lab}=8.95$ GeV/c.}\label{1295}
\end{figure}

Another notable behavior for these calculated lines shape is that there exists a small cusp near the production threshold for all of the Regge model results.
They are relevant to the factor $1+\xi e^{-i\pi \alpha(t)}$, which may result in dips at $\alpha_{a_{0}}(t)=-1,-3,-5,\cdots$ for the pion-proton scattering reaction. And, for the kaon-proton scattering reaction, dips appear at $\alpha_{K^*}(t)=0,-2,-4,\cdots$.
These cusps also exist the $X(2100)$ and $h_1(1965)$ productions by the pion-proton scattering and the kaon-proton scattering ~\cite{Wang:2019qyy,Kong:2018bst,Huang:2008nr}.
Whether such cusps are physical or only unphysical should be clarified by the future precise experimental data.

In fact, the above study gives the prediction of searching for higher radial excitations of pseudoscalar meson via the pion-proton scattering and the kaon-pion scattering, which can be a new task for future experiments. Although the present experimental information is still absent, we try to test the validity of our framework adopted in this work. We notice that there exists
measurement of the differential cross section of $\pi^{-}p\rightarrow \eta(1295)n$ at $p_{Lab}=8.95$ $\text{GeV/c}$ \cite{Fukui:1991ps}. Thus, we calculate the differential cross section of $\pi^{-}p\rightarrow \eta(1295)n$ at $p_{Lab}=8.95$ $\text{GeV/c}$ and make a comparison of our theoretical result with the experimental data (see Fig. \ref{1295} for the details). Here,
 the blue doted line and red solid line are calculated in the Feynman model and the Regge model, respectively, and the experimental data from the left-hand side to right-hand side are average values of differential cross section in the range of $|t-t_{min}|$ with $0\sim0.05$ $\text{(GeV/c)}$$^2$, $0.05\sim0.2$ $\text{(GeV/c)}$$^2$ and $0.2\sim0.6$ $\text{(GeV/c)}$$^2$.
Generally, our result is comparable with the experimental data.
In the small $|t-t_{min}|$ range, the Regge model result is closer to the experimental result than the Feynman model result. By this study combined with a concrete experiment, we have reason to believe that the adopted theoretical framework in the present work can be used to
study the production of higher radial excitation in the pseudoscalar meson family, and give reasonable theoretical prediction accessible at experiment.

{As shown in Fig. \ref{2627}, $\eta(2627)$ mainly decays into $\pi a_0(1450)$ and $\pi a_2(1320)$, which means that there should exists the $\pi^-p\to \eta(2627)n$ process via exchanging $a_0(1450)$ and $a_2(1320)$. In our realistic calculation, we only take $a_0(980)$ exchange. A main reason why we do not include the $a_0(1450)$ and $a_2(1320)$ exchange contribution process to $\pi^-p\to \eta(2627)n$ is the absent information of the $NNa_0(1450)$ and $NN a_2(1320)$ coupling, where $N$ denotes nucleon. Thus, this treatment means that the cross sections estimated in this work may not be precise. For the discussed $\eta(2867)$ production, we also face the same situation. }

{Additionally, for $\eta(2742)$ as $\eta^\prime(6S)$, the decay width of its $a_0(980)\pi$ channel is not sizable, which shows that the cross section of $\pi^-p\to \eta(2742)n$ process should be suppressed compared with that $\pi^-p\to \eta(2627)n$. Indeed, the results presented in Figs. \ref{2627s} and \ref{2742s} support this scenario.}

\section{summary}

The study of hadron spectroscopy may provide valuable hints to understand the nonperturbative behavior of QCD. As an important group in whole hadron family, light hadron has attracted extensive attention of theorists and experimentalists. In this work, we still pay attention to the pseudoscalar meson family. Since 2003, more and more pseudoscalar states including $X(1835)$, $X(2120)$, $X(2370)$ and $X(2500)$ have been reported in experiments \cite{Ablikim:2005um,Ablikim:2010au,Ablikim:2016itz,Ablikim:2016hlu}, and provide a  good chance to construct the pseudoscalar meson family \cite{Yu:2011ta,Wang:2017iai}. It is obvious that it is not the end of the whole aspect. Especially, the BESIII measurement of the $\eta^\prime\pi^+\pi^-$ invariant mass spectrum of $J/\psi\to \gamma \eta^\prime\pi^+\pi^-$ \cite{Ablikim:2016itz} shows that there exists a possible event accumulation around 2.6 GeV. This experimental information also stimulates our interest in exploring higher radial excitations of the pseudoscalar meson family.

In the present work, we focus on the fifth and the sixth radial excitations of the pseudoscalar meson family. By performing the mass spectrum analysis and the two-body OZI-allowed strong decay calculation, we may obtain the information of their resonance parameters and partial decay widths, which is crucial for hunting for and identifing these higher radial excitations of the pseudoscalar meson family in future experiments.

Of course, it should not limit us to the above issues since the investigation around
these pseudoscalar mesons contains their production and so on. We also notice that the established $\eta(1295)$ was first observed in the pion-proton scattering process \cite{Stanton:1979ya}. Thus, in this work, we exam the possibility of searching for the discussed four pseudoscalar mesons via the production induced by the pion or kaon. By the effective Lagrangian approach, we estimate the corresponding production cross sections, and find that these physical quantities are sizable. Combining with this  information, we further give a theoretical suggestion of finding them.

As indicated in Ref. \cite{Ablikim:2019hff}, BESIII still plays important role in exploring light hadron. Especially, BESIII has 10 billion $J/\psi$ data, which makes the discovery of these higher radial excited pseudoscalar mesons become possible. Although BESIII has special status of studying on the light hadron, we still believe that it is not a  unique way to detect these states. The pion-proton and kaon-proton scattering experiments can be as a supplement as illustrated in the present work.

\vfil

\section*{Acknowledgments}
This work is supported by the China National Funds for Distinguished Young Scientists under Grant No. 11825503, the National Program for Support of Top-notch Young Professionals, the projects funded by Science and Technology Department of Qinghai Province (Project No. 2020-ZJ-728), and China National Funds under Grant No. 111965016.

\end{document}